\documentstyle[prd,epsfig,aps,floats]{revtex}

\newcommand{\beq}{\begin{equation}}
\newcommand{\eeq}{\end{equation}}
\newcommand{\beqa}{\begin{eqnarray}}
\newcommand{\eeqa}{\end{eqnarray}}

\newcommand{\lsim}{\lesssim}

\newcommand{\ie}{{\it ie.}}
\newcommand{\eg}{{\it eg.}}
\newcommand{\cf}{{\it cf.}}
\newcommand{\etal}{{\it et al.}}

\newcommand{\jpsi}{J/\psi}

\newcommand{\morder}[1]{{\cal O}(#1)}
\newcommand{\eq}[1]{Eq.\ (\ref{#1})}

\newcommand{\rvec}{\bbox{r}}
\newcommand{\pvec}{\bbox{p}}

\newcommand{\as}{\alpha_s}

\newcommand{\cpair}{c\bar c}

\newcommand{\PL}[3]{Phys.\ Lett.\ {{\bf#1}}, {#2} ({#3})}
\newcommand{\NP}[3]{Nucl.\ Phys.\ {{\bf#1}}, {#2} ({#3})}
\newcommand{\PR}[3]{Phys.\ Rev.\  {{\bf#1}}, {#2} ({#3})}
\newcommand{\PRL}[3]{Phys.\ Rev.\ Lett.\ {{\bf#1}}, {#2} ({#3})}
\newcommand{\ZP}[3]{Z. Phys.\ {{\bf#1}}, {#2} ({#3})}

\begin{document}

\twocolumn[\hsize\textwidth\columnwidth\hsize\csname @twocolumnfalse\endcsname
\title{%
\hbox to\hsize{\normalsize\hfil\rm NORDITA-1999/59 HE}
\hbox to\hsize{\normalsize\hfil\rm LAPTH-749/99}
\hbox to\hsize{\normalsize\hfil hep-ph/9909519}
\hbox to\hsize{\normalsize\hfil September 27, 1999} %PH \protect\today}
\vskip 40pt
$\psi '$ to $\psi$ Ratio in Diffractive Photoproduction}

\author{Paul Hoyer}
\address{Nordita, Blegdamsvej 17, DK-2100 Copenhagen, Denmark}

\author{St\'ephane Peign\'e}
\address{LAPTH/LAPP, F-74941 Annecy-le-Vieux Cedex, France}

\maketitle

\vskip2.0pc
\begin{abstract}
We evaluate the $\psi'$ to $\jpsi$ ratio in diffractive photoproduction
in a light-cone framework,
using charmonium wave functions extracted from non-relativistic potential
models. Contrary to current belief, we find that the best estimate for the
ratio is a factor 2 to 5 below the data. The measured ratio constrains the
distribution of the $\cpair$ component of the charmonium light-cone wave
function and indicates 
that it is more compact than in potential models. We
predict that the inelastic photoproduction ratio will be bigger than the
elastic one, and will equal that measured in hadroproduction.
\end{abstract}
\vskip2.0pc]

%% Section I %%

\section{Introduction} \label{sec1}

Quarkonium production is a hard process in which the heavy quarks are produced
with limited relative momentum. This kinematic restriction implies that the
standard QCD factorization theorem does not apply, \ie, there is no guarantee
that the cross section can be expressed as a product of universal parton
distributions and a partonic subprocess. Quarkonium production
is thus sensitive to the environment and yields new 
information about
the dynamics of hard processes. The abundant data imposes severe restrictions
on theoretical models (for a discussion see Ref. \cite{hp}).
In particular, the
recent CDF data \cite{cdf99} on charmonium polarization
at large $p_{\perp}$ in $p \bar p$ collisions disagrees with the
color octet model prediction \cite{com}.

The ratio of cross sections for radially excited states is sensitive to the
time-scale of the production process. If all relevant interactions
occur while the quark pair is compact, $r_\perp \sim 1/m_Q$, the direct
production cross section is proportional to $|\Phi(0)|^2$, the square of the
quarkonium wave function at the origin.
Conversely, 
late interactions
depend on the shape of the wave function out to its Bohr radius, $r_\perp \sim
\morder{1/(\as m_Q)}$.
The $\sigma(\psi')/\sigma(\jpsi)$ ratio
is thus a very interesting quantity which may give
clues on the correct production mechanism. 

In forward charmonium hadroproduction the ratio of $\psi'$ to $\jpsi$ cross
sections is found to be roughly independent of the kinematics
and of the size of the nuclear target \cite{lourenco,e866}. Its value is
moreover consistent with the expectation
based on the proportionality to $|\Phi(0)|^2$ \cite{vhbt},
\beq
R^{hN} \equiv \frac{\sigma^{hN}(\psi')}{\sigma^{hN}_{dir}(\jpsi)} =
\frac{\Gamma(\psi' \to e^+e^-)} {\Gamma(\jpsi \to e^+e^-)}
\frac{M^3_{\jpsi}}{M^3_{\psi'}} \simeq 0.24 \>,
\label{ratiohadro}
\eeq
suggesting that late interactions are unimportant for quarkonia
produced in hadron fragmentation regions. 
In nuclear fragmentation regions the ratio is measured to be smaller
than in \eq{ratiohadro}. This is explained by the larger break-up cross section
of the $\psi'$ in late interactions with the nuclear comovers.

Preliminary CDF results \cite{cdf99} show that $52 \pm 12 \%$ of the
$\Upsilon(1S)$ are directly produced. Given that the $\Upsilon(3S)$ is
produced only directly
one deduces from the published total cross sections
\cite{cdf95} that $\sigma(3S)/\sigma_{dir}(1S) = 0.4 \pm 0.1$, consistent
with the expectation $0.3 \pm 0.05$ based on \eq{ratiohadro} for bottomonium.

Diffractive charmonium electroproduction, $\gamma^{(*)}p \to \psi(nS) p$, is
believed to occur via two-gluon exchange \cite{ryskin}.
At large $Q^2$, the size of the quark pair in the virtual
photon wave function is 
$\sim 1/Q \ll 1/(\as m_c)$, and
the cross section is predicted to be proportional
to $|\Phi_n(0)|^2$. The $\psi'$ to $\jpsi$ ratio at
large $Q$ is indeed measured \cite{electro} to be consistent with
$|\Phi_{\psi'}(0)/\Phi_{\jpsi}(0)|^2 \sim 0.5-0.6$ \cite{eiqu}.
On the other hand, for photoproduction \cite{photo} the measured value
\beq
R^{\gamma N}_{el} \equiv\frac{\sigma(\gamma p \to \psi' p)}{\sigma(\gamma p \to
\jpsi p)} = 0.15 \pm 0.034 \ \ \ (Q^2 = 0)
\label{ratiophoto}
\eeq
is about a factor 3 below the value found at large $Q^2$.

It was pointed out \cite{knnz} that due to the moderate value of the charm
quark mass and a factor $r_\perp^2$ from the coupling of the two gluons, the
photoproduction amplitude in fact probes the charmonium wave function at a
transverse size $r_\perp$ which is comparable to the Bohr radius. This reduces
the $\psi'$ contribution due to the node in its wave function.
The range of transverse size which is probed decreases when the
virtuality $Q^2$ of the photon increases. It is thus possible to measure the
shape of the charmonium wave function using electroproduction data.

In Ref.\cite{knnz} an estimate of the node effect gave a value
.15 for the ratio (\ref{ratiophoto}).

This value was obtained with harmonic oscillator wave functions for the bound
states and with the weighted photon wave function parametrized as a sum of two
Gaussian functions \cite{nemchik}. We find that the result is very sensitive to
the  parametrization and is thus uncertain to at least 50\%.

We report here a more quantitative calculation of the
photoproduction ratio, in a light-cone framework using charmonium wave
functions obtained from potential models. Surprisingly, the calculated
ratio (\ref{ratiophoto}) turns out to be a factor 2 to 5 below the data. We
discuss the implications of this for the charmonium wave function.

%% Section II %%

\section{Evaluation of the Ratio} \label{sec2}

The measured $\psi'/\jpsi$ photoproduction ratio is consistent with being
independent of the photon energy \cite{photo,inelastic}.
At the H1 energy
the incoming photon fluctuates into the $\cpair$ pair long before
the target. We therefore work in the high energy regime where the
transverse size $r_\perp$ of the ($S$-wave) $\cpair$  pair is frozen
during its interactions  
in the target and distributed according to the photon wave function
\beq
\Phi_{\gamma}(x, \rvec_\perp) \propto \sqrt{x(1-x)}\, K_0(m_c r_\perp) \>,
\label{gammawf}
\eeq
where $x$ is the light-cone momentum fraction carried by the $c$ quark.

Each exchanged gluon couples to the $\cpair$ pair with a strength proportional
to the color dipole length $r_\perp$ (in the usual approximation where the
gluon wavelengths are long compared to $r_\perp$). The forward scattering
amplitude is then given \cite{BL} by an overlap with the charmonium wave
function $\Phi_{\psi}$:
\beq
{\cal M}_{\psi} \propto \int \frac{dx}{\sqrt{x(1-x)}}
d^2\rvec_\perp \Phi_{\gamma}(x, \rvec_\perp)
r_\perp^2  \Phi_{\psi}(x, \rvec_\perp)^* \>.
\label{overlapxrperp}
\eeq

We consider two models for $\Phi_{\psi}$, obtained from the non-relativistic
Buchm\"{u}ller-Tye (BT) and Cornell potentials, respectively \cite{eiqu}. In
the non-relativistic limit there is a simple relation between the light-cone
amplitude $\Phi_{\psi}(x, \rvec_\perp)$ appearing in \eq{overlapxrperp} and
the equal-time wave function $\Phi_{\psi}^{NR}(\rvec)$ given by the
Schr\"odinger equation (see, \eg, Ref. \cite{fks}). In momentum space,
\beq
\Phi_{\psi}(x, \pvec_\perp) = \frac{2 (p^2 + m_c^2)^{3/4}}{(p_\perp^2 +
m_c^2)^{1/2}}  \Phi_{\psi}^{NR}(\pvec) \>,
\label{wfrelation}
\eeq
where
\beq
x = \frac{1}{2} + \frac{p^z}{2 \sqrt{p^2 + m_c^2}}
\label{kinident}
\eeq
and the relative factor is fixed by the normalization conditions.

Using
\beqa
F(\pvec_\perp) &\equiv& -\int \frac{d^2\rvec_\perp}{8\pi} e^{-i\pvec_\perp
\cdot \rvec_\perp} r_\perp^2 K_0(m_c r_\perp) \nonumber \\
 &=& \frac{p_\perp^2 - m_c^2}{(p_\perp^2 + m_c^2)^3} \>,
\label{fdef}
\eeqa
\eq{overlapxrperp} reads in momentum space
\beq
{\cal M}_{\psi} \propto \int d^3\pvec \frac{(p_\perp^2 + m_c^2)^{1/2}}{(p^2
+ m_c^2)^{3/4}} F(\pvec_\perp) \Phi^{NR}_{\psi}(\pvec)^* \>.
\label{overlappspace}
\eeq

We use the Mathematica program of Lucha and Sch\"{o}berl \cite{schoeberl}
to solve the Schr\"{o}dinger equation for the BT and Cornell potentials. Our
results for the cross section ratio (\ref{ratiophoto}),
\beq
R^{\gamma N}_{el} =
\frac{|{\cal M}_{\psi'}|^2}{|{\cal M}_{\jpsi}|^2} \>.
\eeq
are shown in the first line of Table~\ref{tab}. The squared ratio of wave
functions at the origin shown on the second line is the result predicted for
highly virtual photons.

%%%%%%%%%%%%%%%%%%%%%%%%%%%%%%%%%%%%%%%%%%%%%%%%%%%%%%%%%%%%%%%%%%%%%%%%%%%%%%%
\begin{table}[htb]
\begin{tabular}{ccccccccc}
{\LARGE\strut} {\large $R^{\gamma N}_{el}=\sigma_{\psi'}/\sigma_{\jpsi}$}
& \hspace{.2cm} & {\large BT} & \hspace{.2cm} & {\large Cornell} &
\hspace{.2cm} \\
\hline {\LARGE\strut}
{\large \eq{overlappspace}} & \hspace{.2cm} & {\large 0.033} & \hspace{.2cm} &
{\large 0.070} & \hspace{.2cm} \\
\hline {\LARGE\strut}
$\large |\Phi^{NR}_{\psi'}(0)/\Phi^{NR}_{\jpsi}(0)|^2$ &
\hspace{.2cm} & {\large 0.65} &
\hspace{.2cm} & {\large 0.64} & \hspace{.2cm} \\
\hline {\LARGE\strut}
{\large $r_\perp^2 \rightarrow r_\perp$} & \hspace{.2cm} & {\large 0.19} &
\hspace{.2cm} & {\large 0.23} & \hspace{.2cm} \\
\end{tabular}
\caption{The $\psi'$ to $\jpsi$ ratio calculated with the Buchm\"uller-Tye
(BT) and Cornell wave functions \protect\cite{eiqu} for elastic
photoproduction (first line), large $Q^2$ electroproduction (second line),
and inelastic photoproduction (last line).}
\label{tab}
%\vskip2.0pc]
\end{table}
%%%%%%%%%%%%%%%%%%%%%%%%%%%%%%%%%%%%%%%%%%%%%%%%%%%%%%%%%%%%%%%%%%%%%%%%%%%%%%%

%% Section III %%

\section{Discussion} \label{sec3}

\eq{overlapxrperp} gives a cross section ratio which is an order of magnitude
smaller than the ratio of wave functions at the origin.
This means that the photon wave function, weighted by
$r_\perp^2$, probes the charmonium wave functions at relatively large
separations. As seen from Fig. 1, $r_\perp^2 \Phi_\gamma$ in fact gives
similar weights to the $\psi'$ wave function in the regions
below and above the
node, leading to a near cancellation in the $\psi'$ integral. This is the
reason for the small value of $R^{\gamma N}_{el}$ and
for its sensitivity to the potential.
%%%%%%%%%%%%%%%%%%%%%%%%%%%%%%%%%%%%%%%%%%%%%%%%%%%%%%%%%%%%%%%
\begin{figure}[t]
\center\leavevmode
\epsfxsize=8cm
\epsfbox{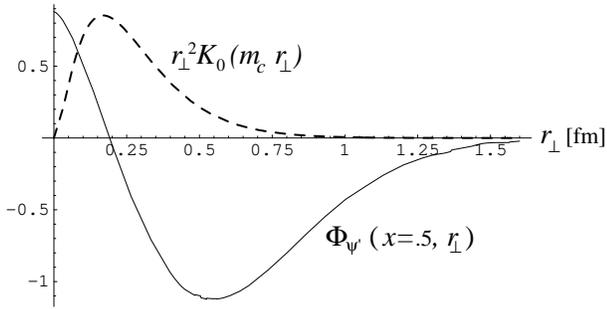}
\vskip 1cm
\caption{The light-cone $\psi'$ wave function $\Phi_{\psi'}(x=.5,r_\perp)$
(solid line) and the weighted photon wave function $r_\perp^2 K_0$ (dashed
line) appearing in the integrand of \eq{overlapxrperp}, plotted versus
$r_\perp$. The normalizations are arbitrary.}
\end{figure}
%%%%%%%%%%%%%%%%%%%%%%%%%%%%%%%%%%%%%%%%%%%%%%%%%%%%%%%%%%%%%%%%%%%%%%%%%

In addition,
the result for the $\psi'$ overlap integral (\ref{overlapxrperp})
is sensitive to the
charmonium wave function in the relativistic domain: only 60\% of the
integral comes from the momentum region $|\pvec| \leq 0.9  m_c$ (the
corresponding number for the $\jpsi$ is 95\%). This means that the theoretical
calculation is not reliable for the $\psi'$ amplitude, since
the wave function was
obtained from the non-relativistic Schr\"odinger equation. The sensitivity to
relativistic momenta has previously been emphasized in Ref. \cite{fks}.

We thus have to conclude that {\em there is no reliable theoretical prediction
for the cross section ratio $R^{\gamma N}_{el}$}.

The weighted photon wave function $r_\perp^2 \Phi_\gamma$ probes $\cpair$
pairs with $r_\perp \lsim 0.5$ fm (\cf\ Fig. 1).
Assuming that \eq{overlapxrperp} is valid in this range implies that
the light-cone charmonium wave functions we used are incorrect,
since our result (Table~\ref{tab}, first line) is a factor 2 to 5 below 
the data (\ref{ratiophoto}). 
On the other hand,
the data may be used to constrain the physical wave functions. In particular,
\begin{itemize}
\item The range of the photon wave function narrows with $Q^2$,
allowing a `scan' of the wave functions \cite{knnz}.

\item The nuclear target $A$-dependence of coherent $\jpsi$ photoproduction is
consistent with $\sigma^{hA} = A^\alpha \sigma^{hN}$, with $\alpha = 4/3$
\cite{sokoloff}. This implies a small rescattering cross section, \ie, a small
transverse size of the $\cpair$ pairs which contribute to $\jpsi$ production.
It would be important to measure the $A$-dependence also in coherent $\psi'$
production.

\item In the case of {\em inelastic} $\jpsi$
photoproduction only one (Coulomb) gluon is exchanged with the target. The
overlap integral corresponding to \eq{overlapxrperp} then
has a {\em single} power
of $r_\perp$. This means that the charmonium wave function is effectively
probed at lower values of $r_\perp$ than in elastic photoproduction.
In our calculation using potential model wave functions the effect of changing
the power of $r_\perp$ in the overlap integral is quite large, as shown in the
last line of Table~\ref{tab}.

\item The mechanism of charmonium hadroproduction is still uncertain, but it
is likely that a single gluon is exchanged with the target. (This is a feature
of all present models \cite{hp,com,cem}).
Just as in inelastic photoproduction, one gluon exchange implies a single
power of $r_\perp$ in the overlap integral (\ref{overlapxrperp}).
This is qualitatively consistent with the measured
hadroproduction ratio (\ref{ratiohadro}), which
is larger than $R^{\gamma N}_{el}$
but smaller than the ratio in large $Q^2$ electroproduction.
\end{itemize}

According to the above arguments,
{\em we expect the
inelastic photoproduction ratio to be the same as that measured in (inelastic)
hadroproduction, $\sigma(\psi')/\sigma_{dir}(\jpsi) \simeq .24$.}   The
preliminary data \cite{inelastic} on inelastic
$\psi'$ photoproduction is still too imprecise for a definite conclusion.

It is perhaps not so surprising that the potential model results for the
charmonium wave functions
are inaccurate. In particular, those models take no account of the
fact that the $\psi'$ lies only 44 MeV below $D\bar D$ threshold. It
follows from the uncertainty principle that the $\psi'$ wave function contains
virtual $D\bar D$ pairs with a lifetime (and size) around 4 fm. The
photoproduction amplitude, however, measures only the $|\cpair\rangle$
component of the wave function. It is quite possible that this component is
narrowly distributed in $r_\perp$, while
$\cpair$ pairs at larger separations are part of higher Fock states which
contain gluons and light quarks.
\eject
\section*{Acknowledgments}
We wish to thank S. J. Brodsky and B. Pire for discussions.
We are grateful for a helpful communication with J. Nemchik, and thank
F. Sch\"{o}berl for sending us his program. We are also grateful for the
hospitality of the theoretical physics group at Ecole Polytechnique, where this
work was initiated. This work was supported in part by the EU/TMR contract EBR
FMRX-CT96-0008.

\end{document}